\documentclass[aps,preprint]{revtex4}
\usepackage[dvips]{graphicx}
\usepackage{amssymb}
\usepackage{amsmath}
\usepackage{color}

\usepackage[normalem]{ulem}
\usepackage{gensymb}
\newcommand{\soutr}{\bgroup\markoverwith{\textcolor{red}{\rule[.5ex]{2pt}{1pt}}}\ULon}

\usepackage{cancel}

\begin{document}
\draft
\title{\bf Investigation of the enhancement factor in the regime of semi-Poisson statistics in a singular microwave cavity}

\author{Ma{\l}gorzata Bia{\l}ous and Leszek Sirko}
\address{Institute of Physics, Polish Academy of Sciences, Aleja Lotnik\'{o}w 32/46, 02-668 Warszawa, Poland}

\date{\today}

\bigskip

\begin{abstract}

We investigated properties of a singular billiard, that is, a quantum billiard which contains a pointlike (zero-range) perturbation. A singular billiard was simulated experimentally by a rectangular microwave flat resonator coupled to microwave power via wire antennas which act as singular scatterers. The departure from regularity was quantitatively estimated by the short-range plasma model in which the parameter $\eta$ takes the values $1$ and $2$ for the Poisson and semi-Poisson statistics, respectively. We show that in the regime of semi-Poisson statistics the experimental power spectrum and the second nearest-neighbor spacing distribution $P(2,s)$ are in good agreement with their theoretical predictions. Furthermore, the measurement of the two-port scattering matrix  allowed us to evaluate experimentally the enhancement factor  $F(\gamma^{tot})$  in the regime of the semi-Poisson statistics as a function of the total absorption factor $\gamma^{tot}$. The experimental results were compared with the analytical formula for $F(\gamma^{tot})$ evaluated in this article. The agreement between the experiment and theory is good.

\end{abstract}
\bigskip
\maketitle

% {\b {Introduction}}
\smallskip

\section{Introduction}

The phenomena of quantum chaos \cite{Wigner1955,Dyson1962,Brody1981} have been intensively investigated during the last decades. The characteristic feature of quantum manifestations of classical chaos are strong state correlations and level repulsion in energy spectra.
The statistical properties of the energy eigenvalues belonging to a complex quantum system can be modeled by the ensembles of random matrices. The random matrix theory (RMT) predicts three main universality classes with the symmetry indices $\beta = 1, 2, \mbox{and } 4$ for orthogonal, unitary and symplectic symmetries, respectively, depending on the symmetries of Hamiltonians \cite{Akemann2011}. The RMT approach was initially developed to study complex scattering properties in nuclear physics \cite{Dyson1963}. By now it has been verified in plenty of experimental, numerical, theoretical studies and applied to many areas, mainly e.g. in nuclear regarding the interactions between particles \cite{Weidenmuller2009,Dietz2017c}, condensed matter \cite{Serbyn2016}, microwave flat billiards \cite{Dietz2010,Zheng2006,Stockmann1990,Sridhar1991,Sirko1997,Hlushchuk2000,Hlushchuk2001, Hlushchuk2001b,Savytskyy2004,HemmadyPRL2005,Hul2005,Dietz2015,BialousPRE2019,Dietz2019} and networks \cite{Hul2004,Hul2012,Dietz2017a,Lawniczak2019,Lawniczak2019b,Lawniczak2021,Lawniczak2021c}. Appropriately the energy-level spectrum experiences all degrees of repulsion typical for full random matrices, passing from Poissonian to Gaussian orthogonal ensemble (GOE), and sequentially through the Gaussian unitary ensemble (GUE) and Gaussian symplectic ensemble (GSE) \cite{Hul2004,Bialous2016b,Stockmann2016}. One should point out that atoms in strong microwave fields \cite{Blumel1991,Jensen1991,Bellerman1992,Sirko1993,Buchleitner1993,SirkoPRL1993,Bayfield1995,
Sirko1995,Sirko1996,Bayfield1999,Sirko2001,Sirko2002,Galagher2016} are also often used in simulations of complex quantum systems with time reversal symmetry.
On the contrary, the integrable quantum systems exhibit uncorrelated energy eigenvalues, where degeneracies are not prohibited \cite{Berry1977}, are described by the Poisson distribution.

While the transition from the integrable to non-integrable system takes place, the degree of spectra correlation increases. In the series of papers Bogomolny el al \cite{Bogomolny1999,Bogomolny2001a,Bogomolny2001b,Bogomolny2002,Bogomolny2004} proposed a plasma model for the semi-Poisson spectral statistics displaying level repulsion without long-range spectral rigidity, i.e. statistics being intermediate between the RMT and Poisson distributions. In the context of short- and long-range correlations the semi-Poisson systems were studied theoretically in Refs. \cite{Bogomolny2001a,Molina2007} and experimentally using  rectangular cavities containing point-like perturbations \cite{Bialous2016}.

The most common measure of spectral regularity is the nearest-neighbor spacing distribution (NNSD) $P(s)$. It indicates the degree of level repulsion hence is utilized to analyze the short-range fluctuation properties. For regular systems with uncorrelated spectra there is no level repulsion and the NNSD is described by the Poisson distribution $P^{Poisson}(s)=\mbox{e}^{-s}$. However, for the semi-Poisson statistics and systems described by the RMT the probability density to find two closely spaced neighboring eigenvalues is given by $P(s)\sim s^{\beta}$ \cite{Bogomolny2004}. It is important to underline that both systems are also paired with exponential decays of the nearest-neighbor spacing distributions: the semi-Poisson in a form $\mbox{e}^{-(\beta+1) s}$ and the RMT as $\mbox{e}^{-c_{\beta} s^2}$, where $c_{\beta}$ depends on the symmetry class of the considered system.
In this article we will analyze a semi-Poisson system with the symmetry index $\beta=1$.

The purpose of this work is to analyze quantitatively the behavior of an integrable system perturbed by  two point-like scatterers. Such a system is simulated by a microwave rectangular flat billiard containing two point-like antennas. We show that in the frequency range $\nu=8-13.5$ GHz the system behaves like the semi-Poisson one. In this frequency range we experimentally evaluated the elastic enhancement factor (EEF) $F(\gamma^{tot})$ which characterizes  scattering processes in the system \cite{Moldauer1961} and is expressed in terms of the two-port scattering matrix $\hat{S}$ \cite{Mitchell2010,Fyodorov2011,Berry1981,Nock2014,Fyodorov2005}
 \begin{equation}
	\label{Eq.1}
	\hat{S}=\left[
	\begin{array}{c c}
		S_{aa} & S_{ab}\\
		S_{ba} & S_{bb}
	\end{array}
	\right].\mbox{}
\end{equation}

 The elastic enhancement factor measures the excess of elastic processes described by the diagonal elements $S_{aa}$ over the inelastic ones described by the off-diagonal element $S_{ab}$ of the scattering matrix $\hat{S}$. The enhancement factor depends on the symmetry class and the total absorption factor  $\gamma^{tot}$ of the system and has been investigated in many chaotic systems, e.g. microwave billiards and networks \cite{Fiachetti2003,Zheng2006,Savin2006,Dietz2010,Kharkov2013,Bialous2020e,Lawniczak2012,Lawniczak2015}. In this article we study the EEF for the semi-Poisson system  in this way filling the gap between the studies performed for regular and chaotic systems. The experimental EEF is compared with the results of an analytical formula for the EEF in the regime of the semi-Poisson statistics proved in this article.

\section{Experiment}

The semi-Poisson system was experimentally realized using a rectangular microwave resonator simulating a quantum billiard, see a schematic view in Fig.~1(a). Billiards are adequate systems for studying quantum chaos as the degree of chaoticity of their classical dynamics depends only on their shape \cite{Stockmann2000,Haake2001}. Moreover, properties of quantum billiards can be studied experimentally using flat or cylindrical microwave resonators \cite{Stockmann1990,Sridhar1991,Sirko1997,Hlushchuk2001}. Such systems offer also the simplest realization of a singular billiard - a quantum billiard which contains a pointlike (zero-range) perturbation \cite{Tudorovskiy2010}. A singular billiard was simulated experimentally by a rectangular microwave flat resonator coupled to microwave power via wire antennas which act as singular scatterers \cite{Bialous2016}. The rectangular cavity was manufactured from brass of the adjustable length $L_{1} =36.5 - 41.5$ cm and fixed width $L_{2}=20.2$ cm. The height of the cavity $d=8$ mm corresponds to the cut-off frequency $\nu_{max}=c/2d \simeq 18.7$ GHz  with $c$ denoting the speed of light in the vacuum. Below the cut-off frequency only the transverse magnetic modes $TM_0$ exist inside the cavity and the Helmholtz equation describing the electromagnetic field in the microwave resonator  and the two-dimensional Schr\"odinger equation for the quantum billiard of the corresponding shape with Dirichlet boundary conditions at the side walls of the resonator are mathematically equivalent. That is why the microwave flat resonators enable to investigate the two-dimensional quantum billiards with respect to a transient region between regular and chaotic dynamics.   The top plate of the resonator contains five randomly distributed holes marked from 1 to 5. The measurement of the two-port scattering matrix $\hat{S}$ was realized by introducing two microwave antennas into the cavity. They perform the role of M=2 scattering channels. We used 3 mm long antennas with the diameter of the wire 0.9 mm in order to minimize the destructive influence on cavity modes.  The two-port scattering matrix $\hat{S}$
 was determined experimentally in the frequency window $8-13.5$ GHz using the vector network analyzer Agilent E8364B which was connected to the microwave cavity via two antennas (see Fig.~1(a)).  In Fig.~1(b) we show an example of the transmission measurement $|S_{15}(\nu)|$ between the antennas positioned at the holes $1$ and $5$ for the frequency range $8\leq \nu \leq 13.5 $ GHz. In order to measure the spectra for different realizations of the cavity the longer side $L_{1}$ of the cavity was increased in 25 steps of 2 mm  from $36.5$ to  $41.5$ cm.
   The cumulative number of eigenstates increases with the frequency as $N(\nu) \thicksim B\nu^{2}$, where $B=(A\pi) / c^2$ and A is the area of resonator. In our previous studies devoted to long-range correlations in the rectangular cavity containing point-like perturbations \cite{Bialous2016}, the cavity with the same width  $L_{2}=20.2$ cm but larger length $L_{1} =45.9-46.7$ cm  was applied. However, in this realization of a singular microwave billiard we used a shorter cavity, with smaller area, to lower the density of eigenstates and to simplify the analysis of the spectra of the cavity.

\section{Experimental results}

\subsection{Spectral statistics}

For the analysis of the spectral statistics, each set of ordered eigenvalues (energy levels) must be converted to a set of normalized spacing, i.e. each sequence must be unfolded in order to eliminate specific properties of the system. The procedure of unfolding is carried out by replacing the resonance frequencies $\nu_{i}$ by the smooth part of the integrated level density, that is given by the Weyl's formula \cite{Stockmann2000}

\begin{equation}
	\label{Eq.2} \epsilon_{i}=N(\nu_{i}).
\end{equation}

This yields dimensionless eigenvalues $\epsilon_{i}$ with mean value unity, $\langle s\rangle=1$ of the spacing $s_{i}=\epsilon_{i+1}-\epsilon_{i}$ between adjacent levels. The Weyl's formula corresponds to polynomial of the second degree that depends on the area and perimeter of the resonator \cite{Stockmann2000}. The nearest-neighbor spacing distribution  is the most common measure of spectral regularity of quantum system which gives information on short-range correlations. The analytical results for the NNSD for regular systems $P^{Poisson}(s)$ and displaying the semi-Poisson statistics $P^{sP}(s)$ are given by the following formulas \cite{Mehta1990,Bogomolny2001a,Bogomolny2004}

\begin{equation}
	\label{Eq.3} P^{Poisson}(s)=\mbox{e}^{-s},
\end{equation}
\begin{equation}
	\label{Eq.4} P^{sP}(s) =4s\mbox{e}^{-2s}.
\end{equation}

The transition between the Poisson and semi-Poisson distributions can be characterized by the parameter $\eta$ \cite{Relano2021},

\begin{equation}
	\label{Eq.5} P(s,\eta)=\frac{\eta^\eta s^{\eta-1} \mbox{e}^{-\eta s}}{\Gamma(\eta)}; \mbox{  }s\geq 0; \mbox{  } \eta\in [1,+\infty],
\end{equation}	
where  $\Gamma(z) = \int_0^\infty dtt^{z-1} \mbox{e}^{-t}$ is the gamma function. The case with $\eta = 1$ corresponds to the  Poisson
distribution while for $\eta=2$  we deal with the semi-Poisson statistics.

 For analyzing of the short- and long-range correlation functions 9224 resonance frequencies for different configurations of the cavity were identified from the measurements of the scattering matrix $\hat{S}$. In Fig.~2 (a) we show the nearest neighbor spacing distribution $P(s)$  (histogram) obtained for $\nu=8-13.5$ GHz. The experimental NNSD is compared to the Poisson (green dotted line), semi-Poisson (red full line) and GOE (blue dash-dotted line) theoretical distributions.  The fit of the formula (\ref{Eq.5}) (black full circles) to the experimental data yields the parameter $\eta=1.972 \pm 0.049$  which is very close to the semi-Poisson distribution, for which $\eta=2$. The inset in Fig.~2(a) shows the integrated level spacing distribution I(s). Also in this case the experimental data (black squares) are close to the theoretical prediction for the semi-Poisson distribution (red full line).

The behavior of a singular microwave billiard was additionally tested using
the second nearest-neighbor spacing distribution $P(2,s)$ and a long-range correlation function - the power spectrum $\langle P^{P}(k)\rangle$.

The second nearest-neighbor spacing distribution $P^{sP}(2,s)$ in a regime of the semi-Poisson statistics  \cite{Bogomolny2001a}
is given by
\begin{equation}
	\label{Eq.6} P^{sP}(2,s) =\frac{8}{3} s^3\mbox{e}^{-2s}.
\end{equation}

In Fig.~2 (c) we show the experimental second nearest-neighbor spacing distribution $P(2,s)$ (histogram) obtained for a singular cavity in a frequency range $8-13.5$ GHz. The experimental results are compared to the theoretical distribution (\ref{Eq.6}) (violet full line). The agreement between them is very good. Just for comparison in Fig.~2 (c) we show also the  Poisson (green dotted line), semi-Poisson (red full line) and GOE (blue dash-dotted line) nearest neighbor spacing distributions, which, as expected, are completely different from the second nearest-neighbor spacing distribution $P(2,s)$ obtained for the singular microwave billiard.

Another statistical measure which can be used for testing the semi-Poisson statistics is a long-range correlation function - the power spectrum of the deviation of the $q$th nearest-neighbor spacing from its mean value $q$, $\delta_q=\epsilon_{q+1}-\epsilon_1-q$ \cite{Relano2002,Faleiro2004}. The power spectrum for a sequence of $N$ levels is given in terms of the Fourier transform from 'time' $q$ to $k$, $s(k)=|\tilde{\delta}_k |^2$, with
\begin{equation}
\label{Eq.7}
\tilde{\delta}_k=\frac{1}{\sqrt{N}}\sum_{q=0}^{N-1} \delta_q\exp\left(-\frac{2\pi ikq}{N}\right).
\end{equation}

It was demonstrated in Refs.~\cite{Relano2002,Faleiro2004} that the power spectrum can be expressed as follows
\begin{equation}
\label{Eq.8}
\langle s(k)\rangle= \frac{1}{4 \pi^2}\left [ \frac{K(k/N)-1}{(k/N)^2} + \frac{K(1-k/N)-1}{(1-k/N)^2}\right ] + \frac{1}{4 \sin^2(\pi k /N)} +\Delta,
\end{equation}

where $\Delta=-1/12$ for the Gaussian ensembles and $\Delta=0$ for Poissonian random numbers. The spectral form factor $K(\tau)$,

\begin{equation}
\label{Eq.9}
K^{Poisson}(\tau) = 1,
\end{equation}
\begin{equation}
\label{Eq.10}
K^{sP}(\tau)=\frac{2+\pi^2 \tau^2}{4+\pi^2 \tau^2},
\end{equation}
\begin{equation}
\label{Eq.11}
K^{GOE}(\tau)= 2\tau -\tau\ln(1+2\tau).
\end{equation}

For $\tilde k = k/N\ll 1$ the power spectrum  exhibits a power law dependence $\langle s(\tilde k)\rangle\propto (\tilde k)^{-\alpha}$. For regular systems $\alpha =2$ and for chaotic ones $\alpha =1$ independently of whether time-reversal invariance is preserved or not. The power spectrum was studied numerically in Ref.~\cite{Robnik2005,Salasnich2005,Santhanam2005,Relano2008,Relano2021} and experimentally in  microwave billiards \cite{Faleiro2006,Bialous2016} and networks \cite{Bialous2016b,Dietz2017a}.

In Fig.~3 we compare the experimental power spectrum $\langle s(k)\rangle$ obtained in the frequency range $\nu=8-13.5$ GHz  (black squares) with the theoretical one $\langle s^{sP}(k)\rangle$ predicted for the semi-Poisson statistics (red full line).  The corresponding results for Poisson and GOE statistics are shown as green dotted and blue dash-dotted lines, respectively. Also here a close agreement of the experimental data with the theoretical prediction for the semi-Poisson statistics is observed.

In conclusion, taking into account our experimental results obtained for the NNSD $P(s)$, the second nearest-neighbor spacing distribution $P(2,s)$ and the power spectrum $\langle s(k)\rangle$, the spectral properties of the rectangular microwave billiards in the frequency range $\nu=8-13.5$ GHz are well described by the short-range plasma model which leads to the so-called semi-Poisson statistics.

\subsection{Elastic enhancement factor}

The elastic enhancement factor $F(\gamma^{tot})$  of the two-port scattering matrix $\hat{S}$ is defined by the following relationship \cite{Fyodorov2005,Savin2006}

\begin{equation}
	\label{Eq.12}
	F(\gamma^{tot})=\frac{\sqrt{\mbox{var}(S_{aa})\mbox{var}(S_{bb})}}{\mbox{var}(S_{ab})},
\end{equation}

where $\mbox{var}(S_{ab}) \equiv \langle |S_{ab}|^2\rangle
-|\langle S_{ab} \rangle |^2$ is the variance of  matrix element $S_{ab}$.

For GOE systems in RMT $2\leq F(\gamma^{tot}) \leq 3 $ and for large $\gamma^{tot}$  the enhancement factor should saturate to $F(\gamma^{tot})=2 $ \cite{Fyodorov2005,Savin2006,Zheng2006,Sokolov2015}. For Poissonian uncorrelated levels $F(\gamma^{tot})=3$.

The elastic enhancement factor $F(\gamma^{tot})$ in the regime of the semi-Poisson statistics, corresponding to the frequency range: $\nu=8-13.5$ GHz, was evaluated using 0.025 GHz sliding window for 150 different realizations of the cavity length and antennas positions. In order to remove significant fluctuations of the EEF, the experimental points were averaged in $(\nu- \delta\nu/2, \nu + \delta\nu/2)$ window, where $\delta\nu=0.5$ GHz.
The total absorption factor $\gamma^{tot}$ of the cavity depends on microwave frequency. In the presence of two open channels (antennas) with the transmission
coefficients $T_a$ and $T_b$, corresponding to an openness $\xi = T_a + T_b$ and
internal absorption $\gamma$, the total absorption width of
the resonances is given by $\gamma^{tot} = 2\pi \Gamma /\Delta= T_a + T_b + \gamma$, where $\Gamma$ and $\Delta$ are the width of resonances and the mean level spacing \cite{Fyodorov2005,Savin2006}. The measurements of the diagonal elements  $S_{aa}$ and $S_{bb}$ of the two-port scattering matrix $\hat{S}$ showed that the transmission coefficients $T_{i} = 1 - |\langle S_{ii}\rangle|^{2}$, where $i=a,b$, are the same. It was found out that the change of the microwave frequency $\nu$ from 8 to 13.5 GHz cased the increase of the total absorption factor  $\gamma^{tot}$ from 1.5 to 4.
Fig.~4(a) shows the experimental results obtained for the elastic enhancement factor $F(\gamma^{tot})$ (black circles). The error bars indicate the standard deviations. In order to show the dependence of $F(\gamma^{tot})$ on both $\nu$ and $\gamma^{tot}$  the upper and lower axes in Fig.~4(a) are labeled by the frequency $\nu$ and the total absorption factor $\gamma^{tot}$, respectively. The two broken lines $F(\gamma^{tot})=3$, $F(\gamma^{tot})=2.5$ show the limits for the semi-Poisson statistics which correspond respectively to very small and very large $\gamma^{tot}$.

Until now there have been no theoretical predictions for the elastic enhancement factor $F^{sP}(\gamma^{tot})$ in the regime of the semi-Poisson statistics. Therefore, in this article we present an
analytical formula for $F^{sP}(\gamma^{tot})$  which allows us to compare the experimental results with the theoretical ones.

The elastic enhancement factor for the symmetry index $\beta=1$ \cite{Savin2006} is defined as

\begin{equation}
	\label{Eq.13}
	F(\gamma^{tot})=2+\delta_{1,\beta=1}-\int^{\infty}_0ds \mbox{e}^{-s}b_{2,\beta=1}\left(\frac{s}{\gamma^{tot}}\right),
\end{equation}
 where $\delta_{i,j}$ is the Kronecker delta. The form factor $b_{2,\beta=1}(\tau)$ is related to the spectral form factor $K(\tau)$ defined by Eqs. (\ref{Eq.10}-\ref{Eq.11})  through the relationship
 \begin{equation}
	\label{Eq.14}
	b_{2,\beta=1}(\tau)=1-K(\tau).
\end{equation}

In particular, for the semi-Poisson statistics \cite{Bogomolny2004}
\begin{equation}
	\label{Eq.15}
b^{sP}_{2,\beta=1}(\tau)=1-K^{sP}(\tau)=\frac{2}{4+\pi^2 \tau^2}\,.
 \end{equation}

 In the limiting cases of very small or very large $\gamma^{tot}$, $b^{sP}_{2,\beta=1}(\infty)=0$ and $b^{sP}_{2,\beta=1}(0)=\frac{1}{2}$, respectively, one obtains

\begin{equation}
	\label{Eq.16}
F^{sP}(\gamma^{tot}) = \begin{cases} 3 & \mbox{at } \gamma^{tot} \ll 1, \\ \frac{5}{2} & \mbox{at } \gamma^{tot} \gg 1. \end{cases}
\end{equation}

The evaluation of the elastic enhancement factor $F^{sP}(\gamma^{tot})$ using the formulas (\ref{Eq.13}) and (\ref{Eq.15})  yields

\begin{equation}
	\label{Eq.17}
F^{sP}(\gamma^{tot}) = 3-\frac{\gamma^{tot}}{\pi}\left[\mbox{ci}(\frac{2\gamma^{tot}}{\pi})\sin(\frac{2\gamma^{tot}}{\pi})-\mbox{si}(\frac{2\gamma^{tot}}{\pi})\cos(\frac{2\gamma^{tot}}{\pi})\right],
\end{equation}

where $\mbox{si}(x)=-\int_x^{\infty}\frac{\sin(t)}{t}dt$ and $\mbox{ci}(x)=-\int_x^{\infty}\frac{\cos(t)}{t}dt$ are the sine and cosine integrals. The integral in Eq.(\ref{Eq.13}) was formally performed using the formula 3.354.1 in Ref. \cite{Gradshteyn2007}.

In Fig~4(a) we compare the experimental results with the theoretical ones predicted by the formula (\ref{Eq.17}) (red full line).  Both experimental and theoretical results are in good agreement.

In Fig~4(b) we show the theoretically predicted dependence of the elastic enhancement factor $F^{sP}(\gamma^{tot})$ on the total absorption factor $\gamma^{tot}$ (red full line). It diminishes gradually from the value 3 at very weak $\gamma^{tot}$ to 2.5 at very large total absorption. The green rectangle shows the frequency range $\nu=8-13.5$ GHz considered in this analysis. The broken line $F(\gamma^{tot})=2.5$ marks the limit for the semi-Poisson statistics which corresponds to very large $\gamma^{tot}$. Just for comparison in Fig~4(b) we also show the elastic enhancement factor $F^{GOE}(\gamma^{tot})$ predicted for GOE systems \cite{Savin2006} (blue dash-dotted line). On contrary to $F^{sP}(\gamma^{tot})$ the elastic enhancement factor $F^{GOE}(\gamma^{tot})$ approaches the value 2 for very large $\gamma^{tot}$ \cite{Savin2006}.

\section{Conclusions}

We evaluated experimentally the elastic enhancement factor  $F(\gamma^{tot})$  in the regime of the semi-Poisson statistics as a function of the total absorption factor $\gamma^{tot}$. In order to compare the experimental results with the theoretical ones we derived the analytical formula for $F^{sP}(\gamma^{tot})$ in this regime. We demonstrated that the agreement between the experiment and theory is good.

\section{Acknowledgments}

This work was supported by the National Science Center, Poland, Grant No 2018/30/Q/ST2/00324.

\pagebreak

%\centerline {\bf Figure Captions}

\begin{figure}[tb]
\begin{center}
\rotatebox{0}{\includegraphics[width=1.4\textwidth,
height=0.6\textheight, keepaspectratio]{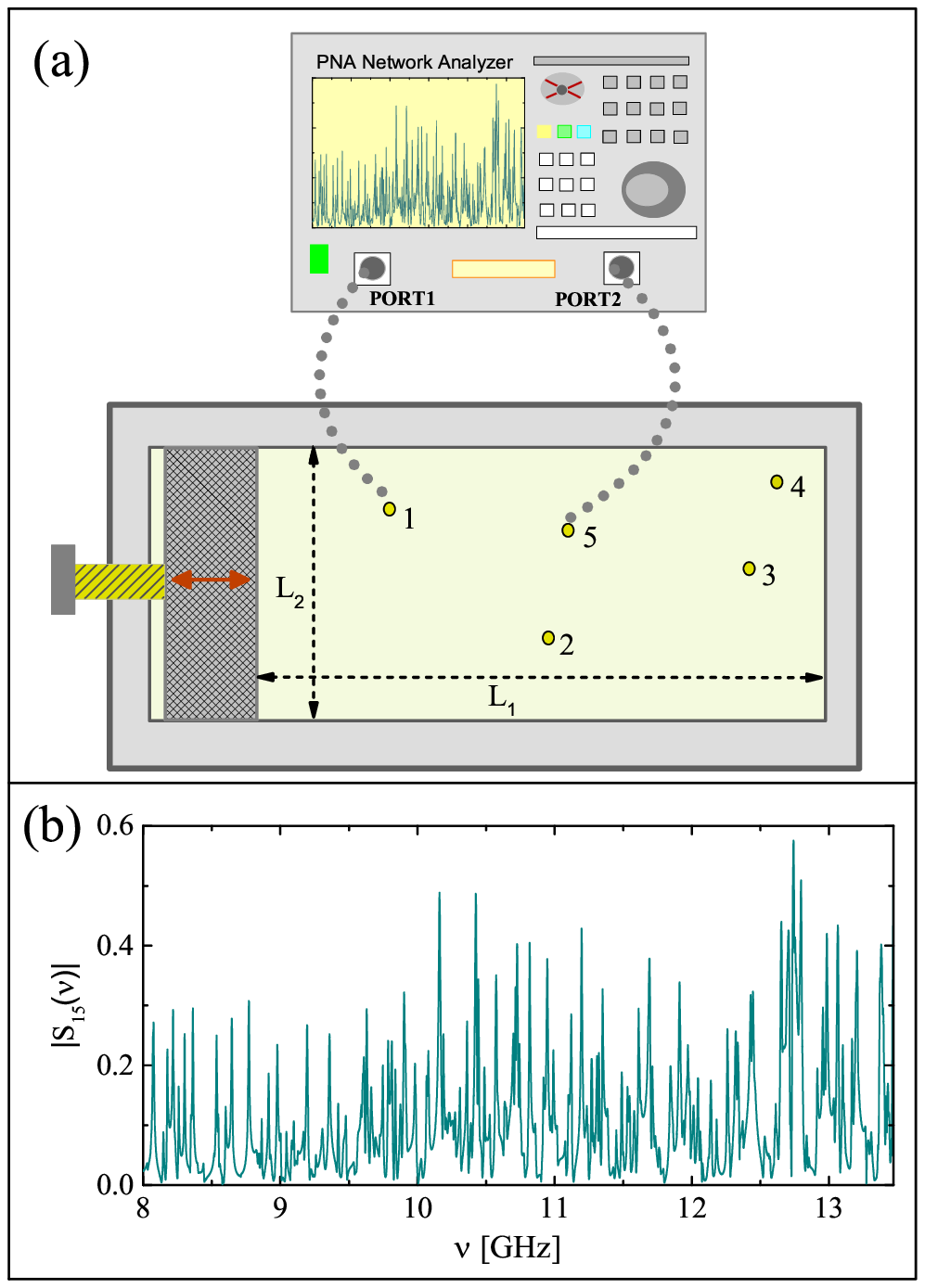}}
\caption{(a) The rectangular microwave cavity of the fixed width $L_{2}$ = 20.2 cm and length $L_{1} = 36.5-41.5$ cm which was changed in 25 steps of 0.2 cm length in order to create different realizations of the cavity. Two microwave antennas introduced inside the resonator (holes 1, 2, 3, 4, 5) allowed to connect the system to the vector network analyzer (VNA) Agilent E8364B through flexible microwave cables HP 85133-616 and HP 85133-617. The two-port scattering matrix $\hat{S}$ of reflected and transmitted spectra was measured in the frequency range $[8-13.5]$ GHz. (b) An example of the transmission measurement $|S_{15}(\nu)|$ between the antennas positioned
at the holes $1$ and $5$ for the frequency range $8\leq \nu \leq 13.5 $ GHz.
}\label{Fig1}
\end{center}
\end{figure}

\begin{figure}[tb]
\begin{center}
\rotatebox{0}{\includegraphics[width=1.2\textwidth,
height=0.9\textheight, keepaspectratio]{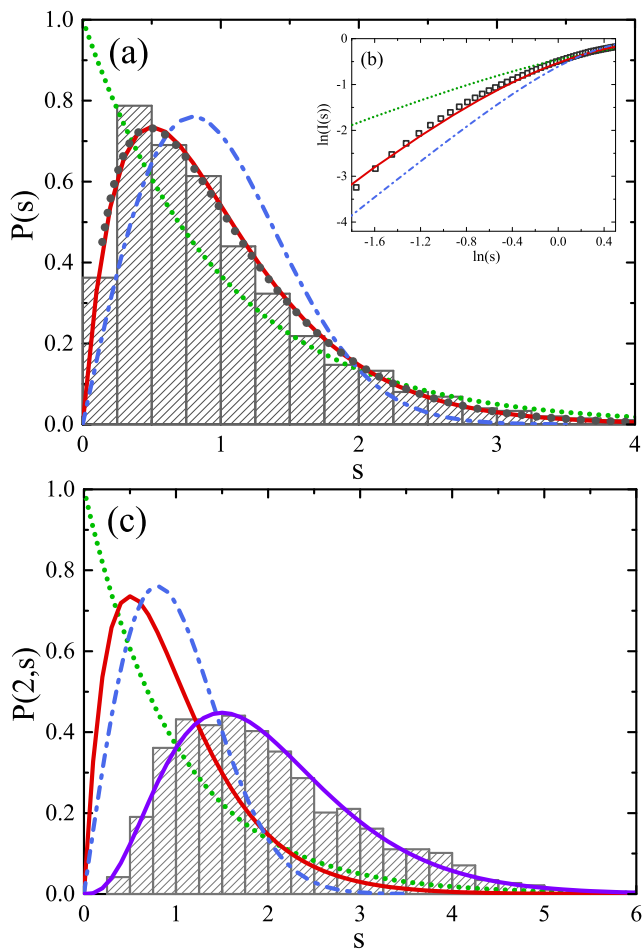}}
\caption{
Nearest neighbor-spacing distribution $P(s)$ (histogram) obtained in the frequency range $\nu=8-13.5$ GHz (panel (a)). The experimental NNSD is compared to the Poisson (green dotted line), semi-Poisson (red full line) and GOE (blue dash-dotted line) theoretical distributions.  The fit of the formula (\ref{Eq.5}) (black full circles) to the experimental data yields the parameter $\eta=1.972 \pm 0.049$  which is very close to the semi-Poisson distribution, for which $\eta=2$. In the inset in Fig.~2 we show the integrated level spacing distribution I(s). The experimental data (black squares) are compared to the theoretical prediction for the semi-Poisson distribution (red full line). Fig.~2 (c) shows the experimental second nearest-neighbor spacing distribution $P(2,s)$ (histogram) obtained for a singular cavity in a frequency range $8-13.5$ GHz. The experimental results are compared to the theoretical distribution (\ref{Eq.6}) (violet full line). For comparison in Fig.~2 (c) we show also the Poisson (green dotted line), semi-Poisson (red full line) and GOE (blue dash-dotted line) nearest neighbor spacing distributions.
}\label{Fig2}
\end{center}
\end{figure}

\begin{figure}[tb]
\begin{center}
\rotatebox{0}{\includegraphics[width=0.8\textwidth,
height=0.9\textheight, keepaspectratio]{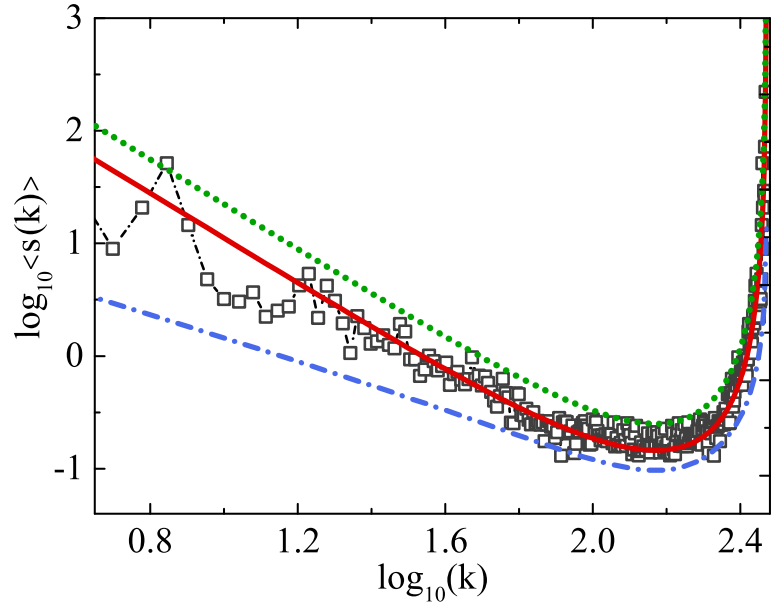}}
\caption{
The experimental power spectrum $\langle s(k)\rangle$ obtained in the frequency range $\nu=8-13.5$ GHz  (black squares) is compared with the theoretical one $\langle s^{sP}(k)\rangle$ predicted for the semi-Poisson statistics (red full line).  The corresponding results for Poisson and GOE statistics are shown as green dotted and blue dash-dotted lines, respectively.
}\label{Fig3}
\end{center}
\end{figure}

\begin{figure}[tb]
\begin{center}
\rotatebox{0}{\includegraphics[width=0.8\textwidth,
height=0.9\textheight, keepaspectratio]{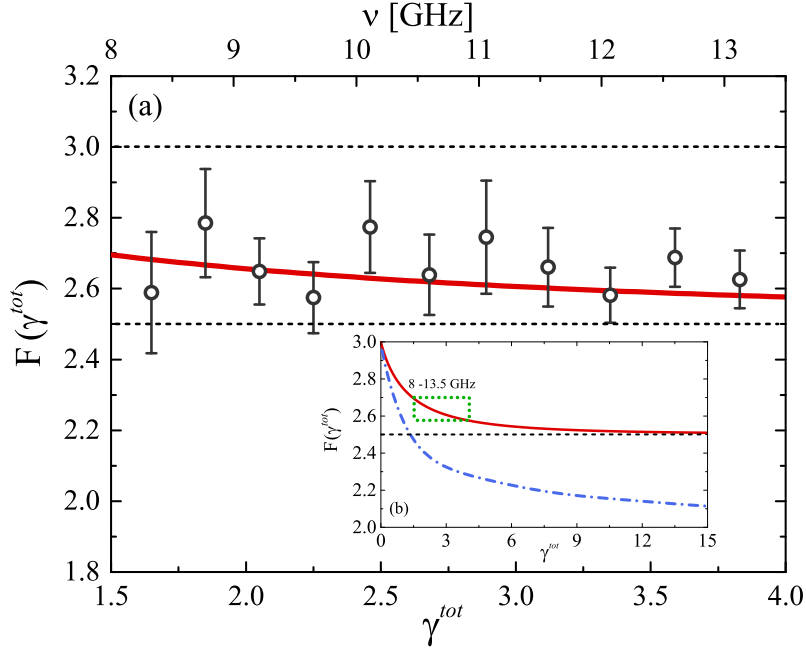}}
\caption{(a)
The experimental elastic enhancement factor $F(\gamma^{tot})$ (black circles). The error bars indicate the standard deviations.  The experimental results are compared with the theoretical ones (red full line). To show the dependence of $F(\gamma^{tot})$ on both $\nu$ and $\gamma^{tot}$  the upper and lower axes are labeled by the frequency $\nu$ and the total absorption factor $\gamma_{tot}$, respectively. The two broken lines $F(\gamma^{tot})=3$, $F(\gamma^{tot})=2.5$ show the limits for the semi-Poisson statistics which correspond respectively to very small and very large $\gamma^{tot}$. (b) The theoretically predicted dependence of the elastic enhancement factor $F^{sP}(\gamma^{tot})$ (red full line) on the total absorption factor $\gamma^{tot}$. The green rectangle shows the frequency range $\nu=8-13.5$ GHz considered in this article.  The broken line $F(\gamma^{tot})=2.5$ marks the limit for the semi-Poisson statistics which corresponds to very large $\gamma^{tot}$. The elastic enhancement factor $F^{GOE}(\gamma^{tot})$ predicted for GOE systems \cite{Savin2006} is shown using the blue dash-dotted line.
}\label{Fig4}
\end{center}
\end{figure}

\end{document}